\documentclass[prl,twocolumn,amsmath,amssymb,showpacs]{revtex4-1}
\usepackage{lipsum}
\usepackage{booktabs}
\usepackage{graphicx}
\usepackage{dcolumn}
\usepackage{bm}
\draft 

\begin{document}


\title{Fabrication of superconducting MgB$_{2}$ thin films by magnetron co-sputtering on (001) MgO substrates} 



\author{Savio Fabretti}
\email[]{fabretti@physik.uni-bielefeld.de}

\author{Patrick Thomas}
\author{Markus Meinert}
\author{Andy Thomas}

\homepage[]{www.spinelectronics.de}
\affiliation{Thin films and physics of nanostructures, Bielefeld University, Germany}


\date{\today}

\begin{abstract}
We fabricated superconducting MgB$_{2}$ thin films on (001) MgO substrates. The samples were prepared by magnetron  rf and dc co-sputtering on  heated substrates. They were annealed ex-situ for one hour at temperatures between 450$^{\circ}$C and 750$^{\circ}$C. We will show that the substrate temperature during the sputtering process and the post annealing temperatures play a crucial role in forming MgB$_{2}$ superconducting thin films. We achieved a critical onset temperature of 27.1\,K for a film thickness of 30\,nm. The crystal structures were measured by x-ray diffraction.
\end{abstract}

\pacs{74.25.F-}
\keywords{MgB$_{2}$, superconductor, thin films}
\maketitle
\section{Introduction}
Many spintronic devices such as those used for spinlogic applications \cite{richter}, are based on magnetic tunnel junctions (MTJ) \cite{prinz}. These devices consist of two ferromagnetic electrodes separated by a thin insulating material, and their resistance depends on the relative orientation of the two magnetizations. Generally, the resistance is lower if the two electrodes are aligned in parallel, and this effect is called tunnel magnetoresistance (TMR). The relative resistance change between parallel and anti-parallel alignment is called the TMR ratio \cite{moodera1995}. A large TMR ratio is desired for MTJs in spintronic devices to separate the high and low state. Some Heusler compounds show a very large TMR ratio due to their predicted high spin polarization of 100\% \cite{deGroot} 
because the effective spin polarization near the Fermi level influences the TMR ratio \cite{Julliere}. A spin polarized tunneling (SPT) measurement offers a reliable method for directly determining the spin polarization\cite{meservey}, which is especially interesting in the case of coherent tunneling through MgO tunnel barriers \cite{Mathon, Butler}. However, almost all of these experiments have been conducted using superconducting aluminum counter electrodes (sometimes with a few percent impurities) \cite{yang}. Other materials show poor properties due to the high atomic number, $Z$, of their constituent(s), which leads to increased ($\propto Z^4$) spin-orbit scattering \cite{abrikosov,abrikosov2}.

Magnesium diboride (MgB$_{2}$) is a compound with a high critical temperature (T$_{c}$) of 39\,K, a simple hexagonal crystal structure, a large coherence length and a high critical current density \cite{Nagamatsu, eom}. These features are desirable for technical applications,  especially for fabrication by sputtering, which makes MgB$_{2}$ an interesting material for electronic devices. Also, magnesium and boron have lower atomic numbers than aluminum, and this makes MgB$_2$ for an excellent candidate for SPT measurements. 

A superconducting film thickness that is smaller than the penetration depth of the applied magnetic field is required for SPT measurements. A MgB$_{2}$ superconducting counter electrode, is a good alternative to aluminum due to its  penetration depth of up to 140\,nm \cite{vinod}. 

In our previous work, SPT measurements of Co$_{2}$FeAl, sputtered on an MgO (001) substrate with a superconducting Al-Si electrode, showed a spin polarized tunneling current with P=55\% \cite{Oli}. 

In this work, we prepared and characterized superconducting thin films of MgB$_{2}$ on MgO for use in spin polarized tunneling experiments. Until now, thin films prepared by co-sputtering were thicker than 100\,nm and deposited on various substrates. The  critical temperatures achieved differ with the preparation method and the choice of substrates \cite{ahn, micunek, saito, lee, vaglio}. 
\section{Sample preparation}
We used $10 \times 10$\,mm$^{2}$ (001) MgO substrates for our layer stack. First, an MgO buffer layer of 5\,nm was deposited from a 4\verb+"+ MgO target by rf-magnetron sputtering to obtain a clean surface. The base pressure of this system was 1$\times 10^{-7}$\,mbar and the argon pressure during sputtering was 2.3$\times 10^{-2}$\,mbar with an applied power of 115\,W. 

Next, the sample was transferred through a high vacuum load lock into a second sputtering chamber with a base pressure of 5$\times 10^{-9}$\,mbar and an argon pressure of 2$\times 10^{-3}$\,mbar during sputtering. The sample was placed on a heated substrate holder, which was rotating at 10\,rpm. The substrate temperature (T$_{S}$) was varied between 272$^{\circ}$C and 310$^{\circ}$C. This growth temperature window was chosen based on the work of van Erven et al.\  They could fabricate MgB$_{2}$ superconducting films at growth temperatures between 200$^{\circ}$C and 300$^{\circ}$C by using molecular beam epitaxy (MBE) \cite{moodera}. The MgB$_{2}$ thin films were deposited by rf-magnetron sputtering of boron and dc-magnetron sputtering of magnesium simultaneously.  Both targets had a diameter of 3\verb+"+. With an applied power of 200\,W, the deposition rate was 0.2\,nm/min for boron at room temperature, while with 30\,W applied, the deposition rate of magnesium was 4.1\,nm/sec. Finally, we evaporated an MgO capping layer of 3\,nm to protect the MgB$_{2}$ film. Otherwise, oxides, such as BO$_{x}$ or MgBO, would form in addition to MgO \cite{singh, klieh1, klieh2}. Furthermore, an MgO capping layer reduces the root-mean-square (rms) roughness of the MgB$_{2}$ film and increases the prerequisite for creating further tunneling junctions \cite{moodera}.

After the deposition process, the samples were ex-situ annealed in a high vacuum furnace with a base pressure of 7$\times 10^{-8}$\,mbar at temperatures between 450$^{\circ}$C and 750$^{\circ}$C for one hour . The crystal structures and the film thicknesses were investigated using x-ray diffraction (XRD) and x-ray reflectometry (XRR), respectively.  Cu-K$_{\alpha}$ radiation was used in all cases. The transport measurements were done in four point geometry in a closed-cycle $^4$He cryostate which was cooled down to 3\,K.

During co-sputtering, the heated substrate is necessary due to the large differences between the sputtering rates of Mg and B. On the one hand, if T$_{s}$ is below 288$^{\circ}$C, too much magnesium covers the substrate surface, and  a very small MgB$_{2}$ peak occurs in the XRD spectra. On the other hand, if T$_{s}$ is above 288$^{\circ}$C, a significant loss of Mg produces an insulating Boron film. Thus, the high Mg vapor pressure was exploited to re-evaporate spare Mg from the surface.
\section{Experiments and Discussion}
\begin{figure}
\begin{centering}
 \includegraphics[width=8cm]{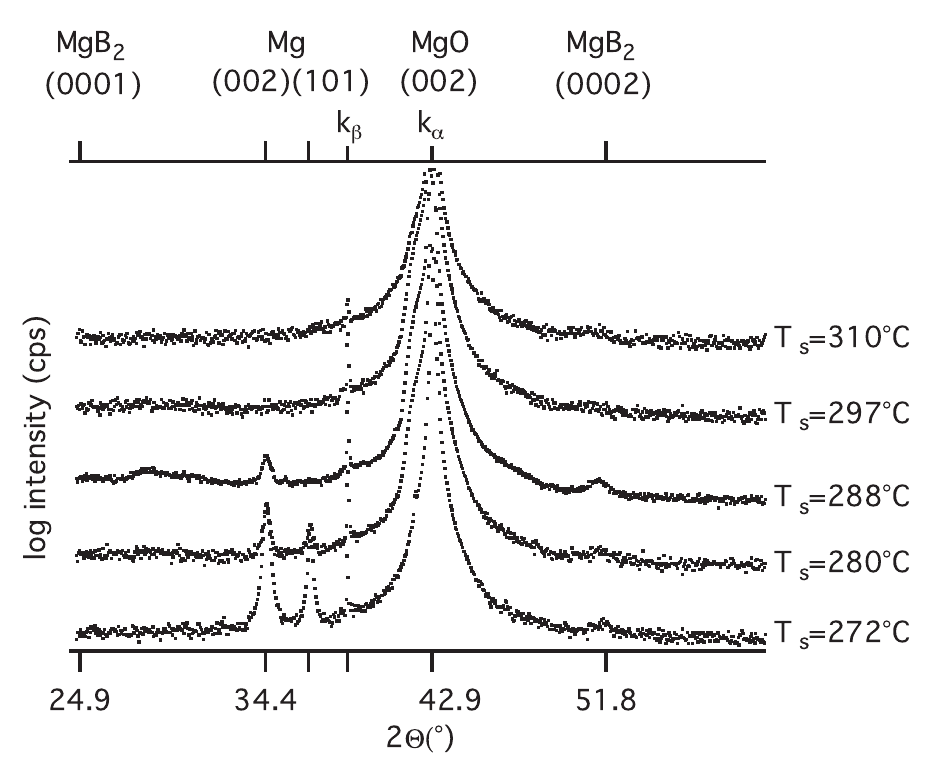}
  \caption{X-ray diffraction scans at different substrate temperatures. T$_{s}$, during deposition. T$_{s}=288^{\circ}C$ shows  weak Mg peaks, and the desired MgB$_2$ (0002) peak is visible.}
  \label{fig:1}
\end{centering}
\end{figure}

First, the predictions about the MgB$_2$ growth were confirmed by $\Theta$-2$\Theta$ x-ray diffraction measurements, shown in Figure \ref{fig:1}, for different substrate temperatures. T$_{s}$, and a fixed MgB$_2$ thickness of $30\pm 5$\,nm.
The (0002) MgB$_{2}$ peak appears between T$_{s}=272^{\circ}$C and T$_{s}=288^{\circ}$C. At low T$_{s}$, crystalline magnesium is observed in the films, but this vanishes at T$_{s}\geq 297^{\circ}$C. Above T$_{s}=288^{\circ}$C, no MgB$_{2}$ peak could be observed. Hence, we chose the sample with T$_{s}=288^{\circ}$C for further investigations because it had the lowest pure Mg content, as shown by the (002) Mg peak, and the clearest (0002) MgB$_{2}$ peak.

Our growth temperature window is slightly different than that  reported by Saito et al.\ \cite{saito}, but comparable to the work of van Erven et al.\ by MBE \cite{moodera}. This discrepancy could be explained by a dependence of T$_{s}$ on the sputtering pressure and the applied sputtering power.
\begin{figure}
\begin{centering}
 \includegraphics[width=8cm]{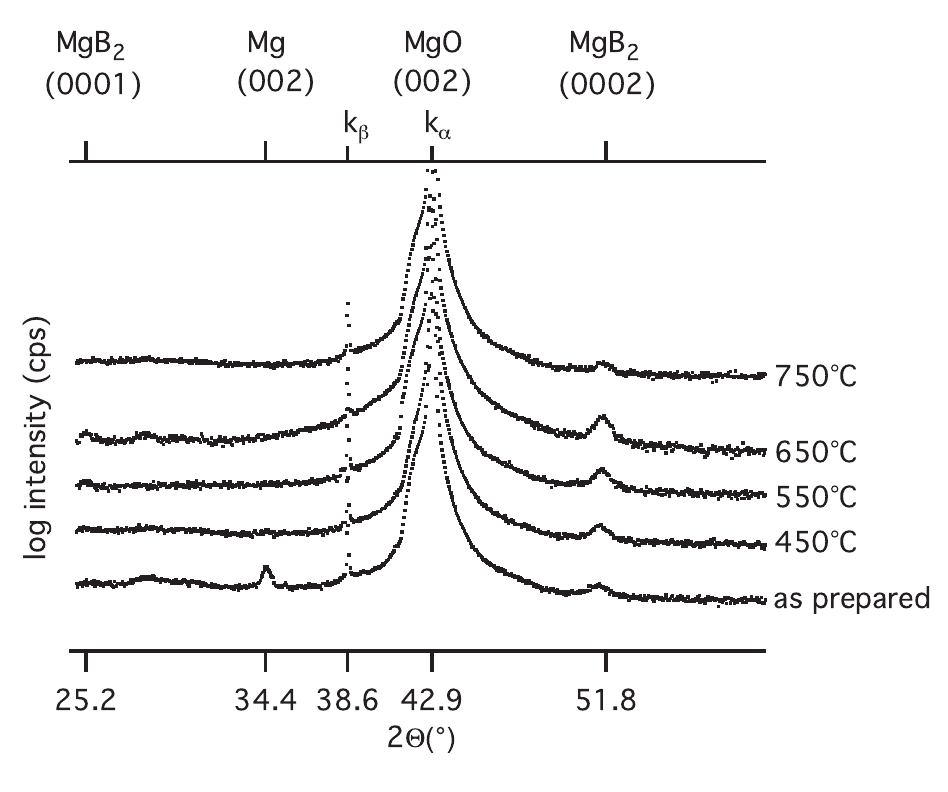}
  \caption{X-ray diffraction scans after one hour post annealing at the given temperatures. The substrate temperature during deposition was always T$_{s}=288^{\circ}C$.}
  \label{fig:2}
\end{centering}
\end{figure}

In Figure \ref{fig:2}, the dependence of the XRD spectra  on the ex-situ annealing temperatures is shown. After the annealing process, the (002) Mg peak has vanished due to the high volatility of magnesium. The (0002) MgB$_{2}$ reflection increases up to an annealing temperature of 650$^{\circ}$C. The film thickness of the samples is 30\,nm. At an annealing temperature of 750$^{\circ}$C, the MgB$_{2}$ peak starts decreasing again.

\begin{table*}
\caption{\label{tab:table3}The x-ray diffraction and transport properties of 30\,nm MgB$_2$ ($T_s=288^\circ C$).}
\label{tab:1}
\begin{ruledtabular}
\begin{tabular}{ccccccc}
 
Annealing temperature&Net Area (arb.u.)&lattice constant c&T$_{c}^{\text{Offset}}$&T$_{c}^{\text{Onset}}$&$\Delta$T$_{c}$&RRR\\ \hline
as prepared&154&3.547&N/A&N/A&N/A&1.134\\
450$^{\circ}$C&159&3.547&18.9\,K&25.8\,K &6.9\,K &1.085\\
550$^{\circ}$C&197&3.538&24.2\,K&27.1\,K&2.9\,K&1.09 \\
650$^{\circ}$C&197&3.536&22.0\,K&25.6\,K&3.6\,K&1.21\\
750$^{\circ}$C&119&3.534&N/A&N/A&N/A&2.4\\
\end{tabular}
\end{ruledtabular}
\end{table*}
 
In Table \ref{tab:1}, the net areas under the (0002) MgB$_{2}$ peaks and the shifting of the out-of-plane lattice constant c are shown. The net areas---the area under the peaks---is roughly proportional to the crystalline volume. A continuous shifting of the (0002) MgB$_{2}$ peak position towards the bulk lattice constant of $c= 3.522$\,{\AA} \cite{wong} was observed. The change of the lattice constant is caused by the mismatch between MgB$_{2}$ and MgO. The  van Erven et al. group reports that the MgB$_{2}$ film is rotated about 45$^{\circ}$ in the plane with respect to MgO. The crystallized MgB$_{2}$ shows a mismatch of $\sim$+3\% for two unit cells on one MgO unit cell \cite{moodera}. In the epitaxial matching case, the MgB$_{2}$ lattice would be compressed in the plane and thus expanded along the c-axis by a few percent. Instead, we observed a lattice parameter approaching the bulk value of MgB$_{2}$. This supports the presumption that the film consists of small, relaxed MgB$_{2}$ grains, which is further supported by the small residual resistivity ratio (RRR) between 1.08 and 1.21, indicating crystallographic defects in the MgB$_{2}$ films. This is in good agreement with Ueda et al.\ who found that a long range crystal ordering is not necessary for superconductivity in MgB$_{2}$ thin films \cite{ueda}. The crystalline volume  increased as annealing temperatures increased from  450$^{\circ}$C to 650$^{\circ}$C. Optimal properties of the lattice structure with respect to the crystalline volume and lattice parameters are observed at 650$^{\circ}$C. The specific resistance is typically between $\rho=100-300\,\mu\Omega$cm at room temperature. This value is comparable with ex-situ post annealed samples \cite{eom}. Due to the loss of Mg at an annealing temperature of 750$^{\circ}$C, the net area of the (0002) MgB$_{2}$ peak is very small. In this case, no superconductivity could be observed. This is confirmed by the relatively high  specific resistance of  $\rho=11300\,\mu\Omega$cm at room temperature. Young Huh's group observed an abrupt  decrease of the Mg content at an annealing temperature of 700$^{\circ}$C \cite{huh}. The sample that was not annealed does not show any indication of superconductivity because it mostly consists of pure magnesium.

\begin{figure}
 \includegraphics[width=8cm]{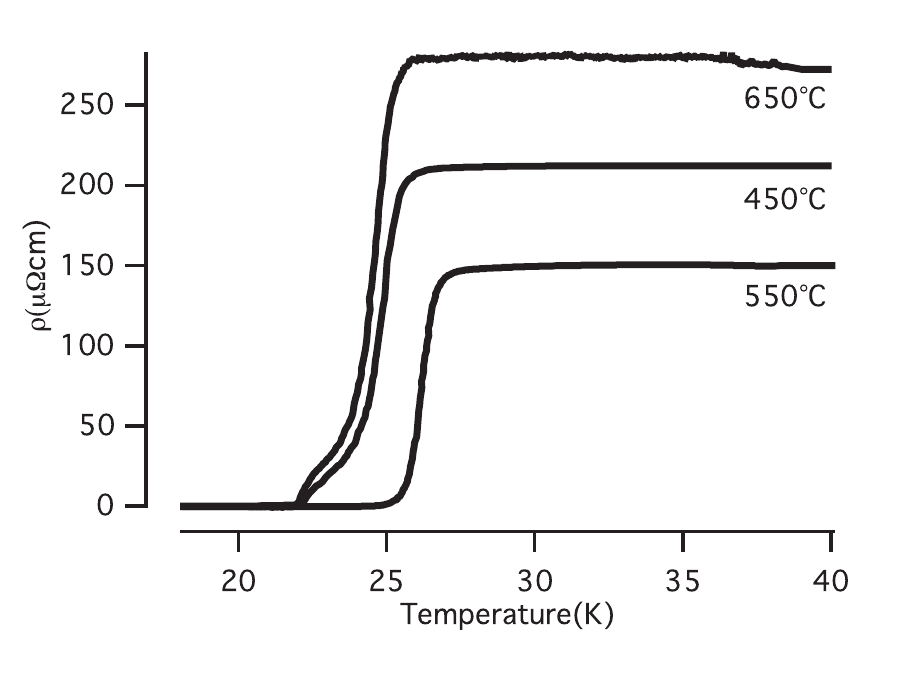}
 \caption{The transport properties of the different samples after one hour post annealing at the  temperatures shown. The substrate temperature during deposition was always T$_{s}=288^{\circ}C$.}
   \label{fig:3}
\end{figure}

In Figure \ref{fig:3}, the transition temperature curves for different annealing temperatures are shown. A dependence on the annealing temperature was previously observed by the group of Mi\v{c}unek \cite{micunek}. They fabricated a 200\,nm thick MgB$_{2}$ film by magnetron co-sputtering. Their samples, sputtered on sapphire substrate, were annealed ex-situ in an argon atmosphere at temperatures ranging  from 500$^{\circ}$C to 700$^{\circ}$C reaching a maximum of T$_{c}$ of 32\,K. Our T$_{c }^{\text{Offset}}$ varied between 18.9\,K and 24.2\,K;  a small enhancement is only visible at an annealing temperature of 550$^{\circ}$C. This sample shows the maximum of T$_{c}^{\text{Onset}}=27.1$\,K. This maximum is in good agreement with the work of Ahn et al.\,, which could reach a T$_{c}^{\text{Offset}}$ of 24\,K at an annealing temperature of 600$^{\circ}$C \cite{ahn}. The sample that was annealed at 550$^{\circ}$C shows the smallest transition width of $\Delta$T$_{c}=2.9$\,K. The relationships between critical temperatures, transition width, RRR  and annealing temperatures are summarized in Table \ref{tab:1}.

In addition to the lattice constant and crystalline volume, the value of $T_{c}$ also depends on the film thickness \cite{ueda}. A film thickness of 30\,nm fabricated with MBE without post annealing resulted in a T$_{c}$ of 29\,K-30\,K \cite{costache}. Furthermore, we suspect that our samples have some contamination with oxygen due to the post annealing process, which can also contribute a reduced T$_{c}$ \cite{singh2}. Additionally, the choice of the substrate has an influence on T$_{c}$. Park et al.\ have shown  that T$_{c}$ is dependent on the substrate. MgB$_{2}$ films sputtered on MgO substrate showed a T$_{c}$ of 24\,K while T$_{c}$ = 37\,K was observed for c-plane Al$_{2}$O$_{3}$ substrates \cite{park}.

Further investigation of the  sputtering parameters of Mg and B and the sputtering pressure will allow us to optimize T$_{c}$, to improve  the transition width and the RRR. However, our relatively high T$_{c}^{\text{Onset}}$ of 27.1\,K shows that we are able to fabricate thin superconducting MgB$_{2}$ films using co-sputtering that are comparable to thin films made by MBE. Using MBE, others have  achieved a  T$_{c}$ of 29-30\,K for a 30\,nm thick MgB$_{2}$ film. \cite{moodera}.
\section{Summary}
We have successfully fabricated superconducting MgB$_{2}$ thin films of 30\,nm thickness on (001) MgO substrates.  During the sputtering process, T$_{s}$, plays an important role in fabricating thin films because it determines their stoichiometry. Low post annealing temperatures of 450$^{\circ}$C showed an onset transition temperature of 26\,K. The highest T$_{c }^{\text{Onset}}$ of 27.1\,K was found for a substrate temperature during deposition of 288$^{\circ}$C and a post annealing temperature of 550$^{\circ}$C. Our films are compatible with magnetron sputtering, a technique employed to  yields the highest TMR ratios of 600\% at room temperature \cite{ikeda}. 
This will allow us to fabricate tunneling junctions for spin polarized tunneling measurements. 
\section{Acknowledgments}
We would like to acknowledge the MIWF of the NRW state government and the German Research Foundation DFG for financial support and we are very grateful to J. S. Moodera and G. Reiss for  encouraging us to start this project.
\bibliography{MgB2}
\end{document}